\documentclass[11pt]{article}
\usepackage{amsfonts} 
\usepackage[utf8]{inputenc}
\usepackage{amssymb, amsmath}
\usepackage{tabularx}
\usepackage{multirow}
\usepackage{amsmath} 
\usepackage{physics}
\usepackage{cite}

\title{The Higgs Field in the closed FLRW universe}

\author{
Metin Arık, Tarik Tok \\
  Department of Physics\\
 Bogazici University\\
   Bebek, Istanbul, Turkey\\
  \texttt{metin.arik@boun.edu.tr}\\
  \texttt{tarik.tok@boun.edu.tr} }

\begin{document}

\maketitle

\begin{abstract}

We consider SO(3) symmetric triplet of Higgs fields and  
SO(4) symmetric complex doublet of Higgs fields in the closed  FLRW universe. For these models, Lagrangian densities provide effective potentials leading to spontaneous symmetry breaking which gives cosmological expectation value of the Higgs field and the Higgs mass. We find a relation which emerges between the size of the FLRW universe and cosmological vacuum expectation value of the Higgs field.
\end{abstract}

\section{Introduction}
Einstein's theory of gravity, also known as General Relativity\cite{rovelli, wald}, is perhaps the most successful and influential scientific theory of modern physics. General Relativity not only explained the anomalous precession of the perihelion of Mercury 
\cite{AEinstein,  preccession}, but also foreshadowed black holes\cite{blackhole, bhole}, gravitational lensing phenomenon\cite{lensing}, gravitational waves\cite{ligo, wave}, and the origin$\And$evolution of the universe\cite{originuniverse}. Einstein showed all these mysterious events with a set of ten coupled, non-linear partial differential equations, called as Einstein field equations \cite{einsteinequation, einsteinfieldequations}. One of the solutions of these equations imply the FLRW metric which assumes that the universe is homogeneous and isotropic \cite{flrwhom}, and it describes the geometry and the dynamics of the universe\cite{flrwstructure, fqflrw}. In other words,  general theory of gravity provides the theoretical foundations for the FLRW cosmology, and the FLRW cosmology allows us to apply Einstein's theory to study the universe as a whole in a simple manner. In section 4, we will discuss the FLRW cosmology at elementary level and then show how to apply it to our model.

Another extraordinary achievement in physics is the Standard Model \cite{Standard}. It explains not only properties of fundamental particles but also their interactions which are the electromagnetic, strong, and weak interactions \cite{electroweak, qcd}. The most important particle of the Standard Model is the Higgs boson. There had been  inconsistency between the theory of particle physics and experimental result of non-zero particle masses before discovery of the Higgs boson by the ATLAS \cite{atlas} and CMS \cite{cms} collaborations at CERN. This uncovered that particles acquired mass through their interactions with the Higgs field. The process through which Higgs field interacts with subatomic particles is named as the Higgs mechanism\cite{kondo}. In this process, the Higgs field breaks symmetry of the electroweak force and the W and Z bosons gain masses \cite{higgsboson, standardmodel11}.

Since ancient times, humans know that the scale of the universe is enormous.  The size of the observable universe is approximately 93 billion light-years at present\cite{diameter}. Cosmology permits us to study the tremendous universe as a whole \cite{wald}. On the contrary, The Standard Model of particle physics does not describe the large-scale structure of the universe directly\cite{peskin}. However, we can conjugate the Standard Model with cosmology to study the behaviour of particles in the formation of the universe \cite{higgsflrw, Arttu}.
Higgs field plays very important roles in the conjunction between the Standard model and the
cosmology\cite{shapos, supergravity, cosmologicalcons}. First of all comes the electroweak phase transition in the evolution of the early universe\cite{phase9, phasetransition}. The Higgs field is closely related to the electroweak phase transition due to the fact that it is responsible for electroweak symmetry breaking and the mass of different essential particles\cite{eweak, eweak2, Electroweaksymmetry}. As the universe evolved from the early hot and dense state to a cold and sparse state, the dynamics of the Higgs field during the electroweak phase transition is relevant for understanding the properties of the universe\cite{eweak3}. Historically, the Higgs field, which gives mass to elementary particles, had been proposed as a possible candidate for the inflaton\cite{inflaton} which causes exponential expansion of the universe soon after the Bing Bang\cite{Inflation}. The final model we would like to mention  is the Higgs portal, which has been proposed as an explanation for dark matter \cite{darkenergy}. It postulates that dark matter particles interact with the Higgs field through a new force\cite{darkmatter}. The dynamics of the Higgs field and its interactions with dark matter could have important implications for the formation of structures in the universe \cite{higgsportal1}. In cosmology, FLRW model is the most useful tool to explain the universe\cite{frwspace, frwspace2}. Not only does its metric describe  a homogeneous and isotropic universe but it  also explains expansion\cite{expansion,  expansion2} or contraction of the universe.  In this paper, we will investigate the connection between FLRW cosmology and the Higgs field.  We would like to consider that, in a dynamic universe which expands in time, the cosmological expectation value of the Higgs field will also be time-dependent\cite{calmet}.

Angular momentum is an important subject in Classical mechanics as well as in General relativity \cite{angular1} and Quantum mechanics\cite{sakurai}. It provides insights into the behavior of rotating systems, the quantization of physical quantities, the behavior of fundamental forces, and the structure and evolution of objects ranging from atoms to galaxies. Despite the fact that scalar fields do not possess intrinsic angular momentum i.e.spin, analogue angular momentum can emerge due to some symmetries in scalar field spaces. In this paper we will take advantage of the angular momentum caused by the SO(3) and SO(4) symmetries to obtain spontaneous symmetry breaking. In quantum theory, analogue angular momentum in scalar field space is quantized, meaning the norm of the Higgs field can only take on discrete values.

Special orthogonal(i.e., SO) groups are useful in various areas of mathematics and physics, such as geometry, mechanics, quantum mechanics, and representation theory\cite{grove}. They provide a mathematical framework for studying rotations and the associated symmetries in different dimensions. We will study spontaneous symmetry breaking in SO(3) and SO(4) symmetric internal scalar field spaces, respectively. The non-Abelian Higgs model with SU(2) gauge symmetry is a crucial component of the Standard Model of particle physics, providing a framework for understanding the electroweak interactions and the origin of particle masses. In this model, the Higgs field acquires a non-zero vacuum expectation value, leading to the spontaneous symmetry breaking of the SU(2) gauge symmetry. This generates masses for the W and Z bosons through the Higgs mechanism. There is an isomorphism between SO(3) and SU(2)\cite{isos}. This shows that the mathematical formalism used to describe rotations and angular momentum in three-dimensional space (SO(3)) is intimately related to the algebraic structure of SU(2). Then, we can consider the non-Abelian Higgs model with SO(3) symmetry as a new model. Thus, the quantum number j associated with the internal SO(3) symmetric scalar field space is conserved and may lead to spontaneous symmetry breaking\cite{angularmomentum}. For the internal SO(4) symmetric space, we will obtain similar results except for the quantization.

In the Standard Model, the Higgs field, by adding a tachyonic negative mass squared term to the Lagrangian density, induces spontaneous symmetry breaking\cite{tachyon}. However, in our model, we will show that spontaneous symmetry breaking can occur regardless of whether we add mass or not. In addition to this, we will make some assumptions to simplify the general solution. The norm of the Higgs field may depend on the size of tachyonic mass, the quantum number associated with the internal symmetries and size of the universe. These dependences are important and leads to more interesting results. The relation between scale factor and the cosmological vacuum expectation of Higgs field will give important clues about in various eras of the FLRW universe.

\section{Space-time metric in FLRW Universe}
The FLRW metric, also known as the Friedmann-Lemaître-Robertson-Walker metric, is a mathematical description of the expanding universe in the framework of general relativity\cite{expansion,  expansion2}.  It provides a way to describe the geometry and dynamics of the universe on large scales.

The square of the infinitesimal line element of FLRW metric is

\begin{equation}
ds^2= dt^2-a^2(t) \frac{d\vec{\zeta}^2}{\big(1+\frac{k\vec{\zeta}^2}{4}\big)^2} 
\end{equation}
where $\vec{\zeta}$ is a three dimensional vector, $a(t)$ is Robertson-Walker scale factor which depends on cosmological time\cite{cosmologicaltime}, t, and k is a dimensionless constant expressing the curvature of the space. The constant can be chosen as -1, 0, 1.

\begin{tabular}{ |p{1cm}||p{8cm}|  }

 \hline
k=-1  & negative curvature (open and curved spacelike section)   \\
  \hline
 k=0&   zero curvature (flat spacelike section)   \\
 \hline 
 k=1 & positive curvature (closed and curved section)\\
 \hline
\end{tabular}

\vskip 0.2in

We will work in a closed universe, so the form of the square of the line element becomes

\begin{equation}
ds^2= dt^2-a^2(t) (dS^3)^2
\end{equation}
where, $dS^3$ is the line element of 3-sphere. Then one can write explicitly as
\begin{equation}\label{metric}
ds^2= dt^2-a^2(t) \big( d\chi^2+sin^2 \chi(d\theta^2+sin^2\theta d\beta^2) \big).
\end{equation}
Here, we choose the convention that dimension of a(t) is $mass^{-1}$ and the angles $\chi$ $\&$ $\theta$ range over $[0,\pi]$ radians and $\beta$ ranges over  $[0,2\pi]$. From (\ref{metric}), the metric tensor of the universe can be written explicitly as

\begin{equation}\label{metric2}
g_{\mu\nu}=  \begin{pmatrix}
   1 & 0 & 0 &0\\[1ex] 
    0 & -a^2  & 0 & 0\\[1ex]
    0 & 0 & -a^2 sin^2\chi & 0\\[1ex]
0 & 0 & 0 & -a^2 sin^2\chi  sin^2\theta   
\end{pmatrix}.
\end{equation}

\section{Vacuum expectation value of Higgs field and angular momentum in internal SO(3) symmetric scalar field space} \label{section3}

SO(3) refers to the special orthogonal group in three dimensions. It describes the rotations in three-dimensional Euclidean space. In the case of a scalar field,  SO(3) field space refers to the field space that transforms under rotations described by the SO(3) group.

In the scalar sector of the Standard Model, there is a single scalar field known as the Higgs field, which is responsible for spontaneous symmetry breaking and the generation of particle masses. The simplest non-Abelian Higgs model is SU(2) symmetric, we consider SO(3) invariance of the field space because of the isomorphism between SU(2) and SO(3) \cite{isos}. 
  The Lagrangian density of the Higgs field can be written as

\begin{equation}\label{lagrangian}
\mathcal{L}= \frac{1}{2}g^{\mu\nu}\partial_\mu \Theta^\dagger \partial_\nu \Theta -V(\Theta)
\end{equation}
with 
$
\Theta=\begin{pmatrix} 
\phi_1 \\
\phi_2 \\
\phi_3
\end{pmatrix}
$
where $\phi_b$ correspond to scalar fields with  $b=1,2,3$, and $\phi_b=\phi^b$. This 3-tuple constitutes SO(3) internal symmetric scalar field space.   Furthermore, the potential term is usually  taken as

\begin{equation}\label{potential}
V(\Theta)=-\frac{1}{2}\bar{m}^2\Theta^\dagger\Theta+\frac{\lambda}{4}(\Theta^\dagger\Theta)^2 
\end{equation}
where $\lambda$ is positive and dimensionless, and  $\bar{m}$ is an absolute value of tachyonic mass. Note that dimension of scalar field is also mass. In general, $\phi_b$ depends on time and the three polar angles in FLRW universe

$$\phi_b=\phi_b(t,\chi,\theta,\beta).$$

In terms of the fields, the Lagrangian density can be written as
\begin{equation} \label{lagden}
\mathcal{L}= \frac{1}{2}g^{\mu\nu}\partial_\mu\phi_b\partial_\nu\phi_b+ \frac{1}{2} \bar{m}^2 \phi_b\phi_b-\frac{\lambda}{4} (\phi_b\phi_b)^2.
\end{equation}

Generally, action of our system may be expressed as

\begin{equation} 
S=\int{}{}\mathcal{L} \sqrt{- g}    dt d\chi d\theta d\beta
\end{equation}
where $g$ is determinant of the metric tensor (\ref{metric2})

\begin{equation}
g=det (g_{\mu\nu})=a^3(t) \sin^2\chi \sin\theta.
\end{equation}

The system can be quantized on time-dependent background by imposing the commutation relation

\begin{equation} \label{commutation}
\left[ \phi_b(t,\chi,\theta, \beta),\pi^a(t,\chi,\theta, \beta)\right]= i \delta(\chi-\chi')\delta(\theta-\theta')\delta(\beta-\beta')\delta_b^a
\end{equation}
where, $\pi_b$ is the canonical momentum density. In FLRW Universe, g is different from 1, then the canonical momentum can be expressed as
\begin{equation} \label{canonical momentum}
\pi_b=\sqrt{- g}\frac{\partial\mathcal{L}}{\partial(\frac{\partial\phi_b}{\partial t})}=\sqrt{- g}
\partial_0\phi_b=a(t)^3 \sin^2\chi \sin\theta  \partial_0\phi_b
\end{equation}

After using (\ref{canonical momentum}), the commutation relation (\ref{commutation}) may be rewritten as

\begin{equation} \label{anglerelation}
\left[ \phi_b(t,\chi,\theta, \beta), \partial_0\phi^a(t,\chi,\theta, \beta)\right]a^3 \sin^2\chi \sin\theta = i \delta(\chi-\chi')\delta(\theta-\theta')\delta(\beta-\beta')\delta_b^a.
\end{equation}

Due to the rotational symmetry of spatial part of FLRW Universe, the isotropic and homogeneous Higgs field depends only on time, $\phi_b=\phi_b(t)$. For the sake of simplicity, we can integrate (\ref{anglerelation}) over all angles

\begin{footnotesize}
\begin{equation} 
\int^{2\pi}_{0}\int^{\pi}_{0}\int^{\pi}_{0}\left[ \phi_a(t), \partial_0\phi^a(t)\right]a^3 \sin^2\chi \sin\theta d\chi d\theta d\beta= i\int^{2\pi}_{0}\int^{\pi}_{0}\int^{\pi}_{0} \delta(\chi-\chi')\delta(\theta-\theta')\delta(\beta-\beta') d\chi d\theta d\beta
\end{equation}
\end{footnotesize}

\begin{equation} \label{com1} 
\left[ \phi_b(t), \partial_0\phi^b(t)\right]2\pi^2 a^3= i 
\end{equation}

One may plug $2\pi^2 a^3$ into (\ref{com1}), and distribute evenly

\begin{equation} \label{com} 
\left[ \phi_b(t)\bigg(2\pi^2 a^3(t) \bigg)^\frac{1}{2}, \partial_0\phi^b(t)\bigg(2\pi^2 a^3(t) \bigg)^\frac{1}{2}\right]= i 
\end{equation}
then define  new fields and their conjugate field momentums

\begin{equation}\label{Phi}
\Phi_b=\bigg(2\pi^2 a^3(t) \bigg)^\frac{1}{2} \phi_b(t)
\end{equation}

\begin{equation}\label{Pi}
\Pi_b=\bigg(2\pi^2 a^3(t) \bigg)^\frac{1}{2}\partial_0\phi^b(t) 
\end{equation}
whose dimensions are $(mass)^{-\frac{1}{2}}$ and $(mass)^{\frac{1}{2}}$ respectively.

Note that, in terms of (\ref{Phi}) and (\ref{Pi}), quantization of the system may be imposed by the quantization relation 

\begin{equation}
\left[ \Phi_b, \Pi^a\right]= i \delta_b^a.
\end{equation}

By definition, the Lagrangian can be expressed as

\begin{equation}
L= \int^{2\pi}_{0}\int^{\pi}_{0}\int^{\pi}_{0} \mathcal{L} \sqrt{- g} d\chi d\theta d\beta.
\end{equation}
We integrate rotational symmetric Lagrangian density (\ref{lagden}) to obtain the Lagrangian

\begin{equation}\label{eql}
L=  \frac{1}{2}2 \pi^2 a^3 \partial_0\phi_b(t) \partial_0\phi_b(t)+\frac{1}{2}2 \pi^2 a^3 \bar{m}^2 \phi_b(t)\phi_b(t)-\frac{\lambda}{4}2\pi^2a^3 (\phi_b(t)\phi_b(t))^2.
\end{equation}
After using (\ref{Phi})\&(\ref{Pi}) and doing some calculations, the Hamiltonian becomes

\begin{equation} \label{hamiltonsec3}
H= \frac{1}{2}\Pi_b \Pi_b-\frac{1}{2} \bar{m}^2 \Phi_b\Phi_b+\frac{\lambda}{4}\big(2\pi^2 a^3\big)^{-1} (\Phi_b\Phi_b)^2.  
\end{equation}

Note that the new field, $\Phi_b$, has internal SO(3) symmetry just like the Higgs fields. Therefore, taking into account of this, we can rewrite the Hamiltonian

\begin{equation}
H=\frac{1}{2}\dot{\Phi}^2+\frac{J^2}{2 \Phi^2}-\frac{1}{2}\bar{m}^2\Phi^2 +
\frac{\Lambda}{4} \big(2\pi^2 a^3\big)^{-1} \Phi^4
\end{equation}
where, $\Phi$ is the norm of $\Phi_b$, $\Phi=\sqrt{\Phi_1^2+\Phi_2^2+\Phi_3^2}$, and $J$ is the dimensionless analogue angular momentum operator of the internal field field space in the view of the SO(3) symmetric conserved quantity. Here, the effective potential is given by

\begin{equation}\label{effective}
V_{eff}=\frac{\mathsf{j}(\mathsf{j}+1)}{2 \Phi^2}-\frac{1}{2}\bar{m}^2\Phi^2 +\frac{\lambda}{4}\big(2\pi^2 a^3\big)^{-1}\Phi^4
\end{equation}
where the eigenvalue of $J^2$ is $\mathsf{j}(\mathsf{j}+1)$, and its minimum is given by the real positive root of 
sixth degree equation

\begin{scriptsize}
\begin{equation} \label{zeqn}
\begin{split}
&\Phi_{min}=\frac{1}{\sqrt{6}}\bigg(\frac{4\bar{m}^2
\pi^2 a^3}{\lambda} \\
&+\frac{2^{\frac{7}{3}}\bar{m}^4 \pi^2 a^3}{\lambda\big(27\lambda^2(2\pi^2 a^3)^{-2}j(j+1)+2\bar{m}^6+3\sqrt{3}\sqrt{\lambda^2(2\pi^2 a^3)^{-2}j(j+1)\big(27\lambda^2(2\pi^2 a^3)^{-2}j(j+1)+4\bar{m}^6\big)^\frac{1}{3}}}  \\
&+\frac{2^{\frac{5}{3}}\big(27\lambda^2(2\pi^2 a^3)^{-2}j(j+1)+2\bar{m}^6}{\lambda} \\
&+\frac{3\sqrt{3}\sqrt{\lambda^2(2\pi^2 a^3)^{-2}j(j+1)\big(27\lambda^2(2\pi^2 a^3)^{-2}j(j+1)+4\bar{m}^6\big)^\frac{1}{3}}
}{\lambda}\bigg)^\frac{1}{2}
\end{split}.
\end{equation}
\end{scriptsize}

Therefore,  the corresponding norm of the Higgs field (\ref{Phi}) can be expressed as

\begin{scriptsize}
\begin{equation} \label{meqn}
\begin{split}
&|\phi_{min}|=\frac{1}{\sqrt{6}}\bigg(\frac{2\bar{m}^2
}{\lambda} \\
&+\frac{2^{\frac{5}{3}}\bar{m}^4}{\lambda\big(27\lambda^2(2\pi^2 a^3)^{-2}j(j+1)+2\bar{m}^6+3\sqrt{3}\sqrt{\lambda^2(2\pi^2 a^3)^{-2}j(j+1)\big(27\lambda^2(2\pi^2 a^3)^{-2}j(j+1)+4\bar{m}^6\big)^\frac{1}{3}}}  \\
&+\frac{2^{\frac{2}{3}}27\lambda^2(2\pi^2 a^3)^{-2}j(j+1)+2\bar{m}^6}{\lambda\pi^2a^3} \\
&+\frac{2^{\frac{2}{3}}3\sqrt{3}\sqrt{\lambda^2(2\pi^2 a^3)^{-2}j(j+1)\big(27\lambda^2(2\pi^2 a^3)^{-2}j(j+1)+4\bar{m}^6\big)^\frac{1}{3}}
}{\lambda\pi^2a^3}\bigg)^\frac{1}{2}
\end{split}
\end{equation}
\end{scriptsize}
where   $|\phi_{min}|=\sqrt{\phi_1^2+\phi_2^2+\phi_3^2}$. Equation (\ref{meqn}) is too complex to analyze directly, then we can show its meaning by focusing special cases.  

Firstly,  we don't need to add tachyonic mass\cite{higgstacyonic} to the effective potential, (\ref{effective}) for obtaining spontaneous symmetry breaking \cite{Jordan,scalarmode}
\begin{equation}\label{zerompot1}
V_{eff}=\frac{\mathsf{j}(\mathsf{j}+1)}{2 \Phi^2} +\frac{\lambda}{4}\big(2\pi^2 a^3\big)^{-1}\Phi^4.
\end{equation}
Here,  its minimum  is given by

\begin{equation}
\Phi_{min}=\bigg(\frac{\mathsf{j}(\mathsf{j}+1)2\pi^2 a^3}{\lambda} \bigg)^\frac{1}{6}. 
\end{equation}
After using (\ref{Phi}),  one can write the norm of the Higss field

\begin{equation}\label{eqnhiggs1}
|\phi_{min}|=\bigg(\frac {\mathsf{j}(\mathsf{j}+1)}{\lambda 4 \pi^4}\bigg)^\frac{1}{6} a^{-1}.
\end{equation}
 Therefore,  we have come to the conclusion that norm of the Higgs fields is inversely proportional to the size of the universe and increasing with the quantum number of analogue angular momentum operator associated with SO(3) symmetry. According to the universe expanding theories, the angular momentum 
in the scalar field space must increase in order for the norm of the Higgs fields to remain constant.

Another case is that is that quantum number angular momentum in SO(3) symmetric field space, $j$ is zero

\begin{equation}\label{zerolpot1}
V_{eff}=-\frac{1}{2}\bar{m}^2\Phi^2 +\frac{\lambda}{4}\big(2\pi^2 a^3\big)^{-1}\Phi^4.
\end{equation}
Then, corresponding the minimum of the norm of the Higgs fields   can be written as

\begin{equation}\label{eqnhiggs3}
|\phi_{min}|=\bigg(\frac {2\bar{m}^2}{\lambda}\bigg)^\frac{1}{2}  .
\end{equation}
Here, it is the standard result that the minimum of norm of the Higgs field does not depend on the size of universe. On the contrary of previous ones, it is constant.

 Spontaneous symmetry breaking can be obtained not only for (\ref{potential}). We can change the sign of the quadratic term in (\ref{potential})

\begin{equation}\label{potential22}
V(\Theta)=\frac{1}{2}\bar{m}^2\Theta^\dagger\Theta+\frac{\lambda}{4}(\Theta^\dagger\Theta)^2 .
\end{equation}

If we follow same steps between (\ref{lagden}) and (\ref{zeqn}), the norm of the Higgs field is

\begin{scriptsize}
\begin{equation} 
\begin{split}
&|\phi_{min}|=\frac{1}{\sqrt{6}}\bigg(-\frac{2\bar{m}^2
}{\lambda} \\
&+\frac{2^{\frac{5}{3}}\bar{m}^4}{\lambda\big(27\lambda^2(2\pi^2 a^3)^{-2}j(j+1)-2\bar{m}^6+3\sqrt{3}\sqrt{\lambda^2(2\pi^2 a^3)^{-2}j(j+1)\big(27\lambda^2(2\pi^2 a^3)^{-2}j(j+1)-4\bar{m}^6\big)^\frac{1}{3}}}  \\
&+\frac{2^{\frac{2}{3}}27\lambda^2(2\pi^2 a^3)^{-2}j(j+1)-2\bar{m}^6}{\lambda\pi^2a^3} \\
&+\frac{2^{\frac{2}{3}}3\sqrt{3}\sqrt{\lambda^2(2\pi^2 a^3)^{-2}j(j+1)\big(27\lambda^2(2\pi^2 a^3)^{-2}j(j+1)-4\bar{m}^6\big)^\frac{1}{3}}
}{\lambda\pi^2a^3}\bigg)^\frac{1}{2}
\end{split}.
\end{equation}
\end{scriptsize}

As in the solution of potential with negative quadratic term (\ref{meqn}), we can examine special cases.

The first is same as (\ref{eqnhiggs1}), because, we don't add tachyonic mass.  Another one  is that the coupling constant, $\lambda$,  is zero,  so that the effective potential can be written as

\begin{equation}\label{effective111}
V_{eff}=\frac{\mathsf{j}(\mathsf{j}+1)}{2 \Phi^2}+
\frac{1}{2}\bar{m}^2\Phi^2 
\end{equation}
where, after some calculations (\ref{Phi}), one can find the minimum of norm of the Higgs field

\begin{equation}\label{eqnhiggs2}
|\phi_{min}|=\bigg(\frac {\mathsf{j}(\mathsf{j}+1)}{4\pi^4\bar{m}^2}\bigg)^\frac{1}{4}  a^{-\frac{3}{2}}.
\end{equation}
Here,  we can make  same comments as for the first case (\ref{eqnhiggs1}), the only difference being the exponents of $a$ and $\mathsf{j}(\mathsf{j}+1)$.

In addition to analyzing special cases, (\ref{eqnhiggs1}), (\ref{eqnhiggs2}), (\ref{eqnhiggs3}) for both potentials (\ref{potential}), (\ref{potential22}), we may make some assumptions to give more details. Now, one can gain a clearer perspective about the behavior of the norm of the Higgs field.

First assumption is

\begin{equation}
j(j+1)\gg|\bar{m}|^6a^{6}
\end{equation}
where, there are two possibilities. Firstly, size of the universe is  too small comparing to quantum number of analogue angular momentum operator. This condition occurs only in the beginning of the universe according to expanding universe theories. Another one is that eigenvalue of analogue angular momentum operator is too big comparing to the size of universe. Although our universe's size is huge, eigenvalue of angular momentum can overwhelm its size, theoretically. In other words, analogue angular momentum of the universe is gigantic in this option. After making this assumption, we obtain

\begin{equation}\label{eqnhiggs4}
|\phi_{min}|\sim \bigg(\pm\frac {4\bar{m}^2}{\lambda}+11 (\bar{m}^2)^2\frac{a^2}{\lambda^{\frac{5}{3}}\mathsf{j}(\mathsf{j}+1)}+ \frac{2}{\lambda^\frac{1}{3}}\frac{(\mathsf{j}(\mathsf{j}+1))^{\frac{1}{3}}}{a^2} \bigg)^\frac{1}{2}  .
\end{equation}
where, ''$+$'' sign refers to (\ref{potential}), and ''$-$'' sign emerges from solution of (\ref{potential22}). The last term in the parentheses is much bigger than others, so the norm of the Higgs field behaves like  this term.

The other is that  analogue angular momentum in scalar field space is not too small comparing to the size of universe

\begin{equation}
|\bar{m}|^6a^{6}\gg j(j+1).
\end{equation}

On the contrary to (\ref{eqnhiggs3},)  anlogue angular momentum in the scalar field space does not have to be zero. It can be in one of too many states, because the universe is known to be very large. Similar to \ref{eqnhiggs3}, the norm of the Higgs fields is the order of tachyonic mass

\begin{equation}
|\phi_{min}|\sim \frac{\bar{m}}{\lambda}.
\end{equation}

Furthermore,  there is an interesting case in these calculations.  The norm of the Higgs filed is the root of  sixth degree equation. The non-numeric solution can be real or complex.  Hovewer,  in certain ranges of variables($m,j,a$),  the complex non-numerical root can be real solution.

In this section, we have seen that many solutions for the norm of the Higgs field depend on the size of universe, (\ref{eqnhiggs1}), (\ref{eqnhiggs2}),  (\ref{eqnhiggs4}). It has different expansion characters
throughout chronology of the universe in the FLRW cosmology.

\section{FLRW cosmolgy}\label{section4}

After General Relativity was developed by Albert Einstein,  Alexander Friedmann and Georges Lemaître independently introduced  expanding universe in the 1920s.\cite{flrwhistory1,  flrwhistory2} Howard P. Robertson and Arthur Geoffrey Walker generalized this idea to maximally symmetric  space-like sections about ten years later.\cite{flrwhistory3,  flrwhistory4} This model based on Einstein's field equations of general relativity

\begin{equation}\label{einsteinfiled}
R_{\mu\nu}-\frac{1}{2}Rg_{\mu\nu}+\Lambda g_{\mu\nu}=8\pi G T_{\mu\nu}
\end{equation}
where,$R_{\mu\nu}$ is Ricci curvature tensor, R is Ricci scalar, $\Lambda$ is cosmological constant, G is the Newtonian constant of gravitation. $T_{\mu\nu}$ is  the stress–energy tensor tensor. There are two independent equations derived Einstein's field equations (\ref{einsteinfiled}) from  describing a spatially homogeneous and isotropic expanding or contracting space-time. They are named as Friedman equations. The first one which derived from 00 componet of equation(\ref{einsteinfiled}) is

\begin{equation}\label{H0}
\frac{H^2}{H^2_0}=\Omega_r \bigg(\frac{a}{a_0}\bigg)^{-4}+\Omega_m \bigg(\frac{a}{a_0}\bigg)^{-3}+\Omega_c \bigg(\frac{a}{a_0}\bigg)^{-2}+\Omega_\Lambda \bigg(\frac{a}{a_0}\bigg)^{-1}
\end{equation}
where, $H_0$ is Hubble parameter at the present time, and the current size of universe is $a_0$. In addition to this, $\Omega_m$ is the matter density, $\Omega_r$ is the  radiation density, $\Omega_c$ the spatial curvature density and $\Omega_\Lambda$ is the cosmological constant(vacuum) density today. Note that 
Hubble parameter may be defined in terms of cosmic scale factor

\begin{equation}\label{H}
H=\frac{\dot{a}}{a}.
\end{equation}

In our model, there is a relation between norm of the Higgs field and eras of the universe due to its cosmic scale factor dependence. From The Big Bang,  the universe  had different the effective energy densities of radiation and matter scale which has to satisfy

\begin{equation}
\Omega_m+\Omega_c+\Omega_r+\Omega_\Lambda=1.
\end{equation}

The size of universe has different expansion characters
throughout chronology of the universe in the FLRW cosmology.\cite{dominatedcosmology}

\subsection{Radiation dominated Universe}

Radiation dominated era began after inflation and taken until $98 \times10^8$ years after the Big Bang. In this period,  radiation which consist of relativistic particles set the dynamics of the universe. Corresponding energy densities were

\begin{equation}
\Omega_r= 1
\end{equation}

 and

\begin{equation}
\Omega_m=\Omega_r=\Omega_\Lambda=0
\end{equation}

After using (\ref{H0}) and (\ref{H}), the scale factor can be found as
\begin{equation}
a(t)=\sqrt{t(t_1-t)}
\end{equation}
where, $t_1$ is the cosmological time of big crunch, and 
we can consider only pre-big crunch era for the real solution of the size of the universe. 

The scale factor dependent solutions,  (\ref{eqnhiggs1}), (\ref{eqnhiggs3}), (\ref{eqnhiggs2}), (\ref{eqnhiggs4}),  respectively,  for radiation dominated universe become

\begin{equation}
|\phi_{min}|=\bigg(\frac {\mathsf{j}(\mathsf{j}+1)}{\lambda 4 \pi^4}\bigg)^\frac{1}{6} \bigg(t(t_1-t))\bigg)^{-\frac{1}{2}}
\end{equation}

\begin{equation}
|\phi_{min}|=\bigg(\frac {2\bar{m}^2}{\lambda}\bigg)^\frac{1}{2}  .
\end{equation}

\begin{equation}
|\phi_{min}|=\bigg(\frac {\mathsf{j}(\mathsf{j}+1)}{4\pi^4(-\bar{m}^2)}\bigg)^\frac{1}{4}  (t(t_1-t))^{-3}.
\end{equation}

\begin{equation}
|\phi_{min}|\sim \bigg(\pm\frac {4\bar{m}^2}{\lambda}+11 (\bar{m}^2)^2\frac{(t(t_1-t))}{\lambda^{\frac{5}{3}}\mathsf{j}(\mathsf{j}+1)}+ \frac{2}{\lambda^\frac{1}{3}}\frac{(\mathsf{j}(\mathsf{j}+1))^{\frac{1}{3}}}{(t(t_1-t))} \bigg)^\frac{1}{2}  .
\end{equation}

\subsection{Matter dominated Universe}
Matter dominated era is between $47 \times10^3$  years and  $98\times 10^8$ years after Bing Bang. The energy density of radiation and the vacuum energy density were surpassed by  the energy density of matter. Effective energy densities can be described as

\begin{equation}
 \Omega_m= 1
\end{equation}
and 
 
\begin{equation}
\Omega_c=\Omega_r=\Omega_\Lambda=0.
\end{equation}

In this era, there is no solution for the size of the universe

\subsection{Cosmological constant dominated Universe}

The cosmological constant-dominated era began after the matter-dominated era. This time coincides $98 \times10^8$ years after the Big Bang. Corresponding energy densities are

\begin{equation}
\Omega_\Lambda= 1
\end{equation}
and

\begin{equation}
\Omega_m=\Omega_r=\Omega_\Lambda=0.
\end{equation}
By using (\ref{H0}) and (\ref{H}), the size of the universe  can be written as

\begin{equation}
a(t)=a_m\cosh(\frac{t}{a_m})
\end{equation}
where, $a_m$ is the minimum size of universe.

The norm of the Higgs fields becomes in the internal $SO(3)$ symmetric space, (\ref{eqnhiggs1}), (\ref{eqnhiggs3}), (\ref{eqnhiggs2}),  (\ref{eqnhiggs4}), in the respective manner, may described as

\begin{equation}
|\phi_{min}|=\bigg(\frac {\mathsf{j}(\mathsf{j}+1)}{\lambda 4 \pi^4}\bigg)^\frac{1}{6} \bigg(a_m\cosh(\frac{t}{a_m})\bigg)^{-1}
\end{equation}

\begin{equation}
|\phi_{min}|=\bigg(\frac {2\bar{m}^2}{\lambda}\bigg)^\frac{1}{2}  .
\end{equation}

\begin{equation}
|\phi_{min}|=\bigg(\frac {\mathsf{j}(\mathsf{j}+1)}{4\pi^4(-\bar{m}^2)}\bigg)^\frac{1}{4}  a^{-\frac{3}{2}}_m\cosh^{-\frac{3}{2}}(\frac{t}{a_m})
\end{equation}

\begin{equation}
|\phi_{min}|\sim \bigg(\pm\frac {4\bar{m}^2}{\lambda}+11 (\bar{m}^2)^2\frac{a^2_m\cosh^2(\frac{t}{a_m})}{\lambda^{\frac{5}{3}}\mathsf{j}(\mathsf{j}+1)}+ \frac{2}{\lambda^\frac{1}{3}}\frac{(\mathsf{j}(\mathsf{j}+1))^{\frac{1}{3}}}{a^2_m\cosh^2(\frac{t}{a_m})} \bigg)^\frac{1}{2}  .
\end{equation}

\section{Vacuum expectation value of Higgs Field and the internal SO(4) symmetric scalar field space}

Spontaneous symmetry breaking can be obtained not only in the internal $SO(3)$ symmetric field space but also in the internal SO(4) symmetric field space. Unlike the previous section, the Higgs fields can also be represented as weak isospin doublet with four components.

\begin{equation}
\Theta=\frac{1}{\sqrt{2}}
\begin{pmatrix}
\phi_1+i\phi_2 \\
\phi_3+i\phi_4
\end{pmatrix}
\end{equation}
where $\phi_b$ corresponds to scalar fields and a=1,2,3,4. Additionally, This doublet constitutes SO(4) symmetric scalar field space.

 After following steps between (\ref{lagrangian}) and (\ref{hamiltonsec3}), we obtain 

\begin{equation}
H=\frac{1}{2}\dot{\Phi}^2 -\frac{1}{2}\bar{m}^2\Phi^2+\frac{\mathbf{J}^2}{2 \Phi^2}+
\frac{\Lambda}{4} \Phi^4
 \end{equation}
where, $\Phi$ is the norm of $\Phi_b$, $\Phi=\sqrt{\Phi_1^2+\Phi_2^2+\Phi_3^2+\Phi_4^2}$, and  $\mathbf{J}$ is the analogue angular momentum operator of the internal field space in the aspect of
the conserved quantities associated with $SO(4)$.  Here, the effective potential for this system is given by

\begin{equation}
V_{eff}=\frac{\mathbf{j}(\mathbf{j}+2)}{2 \Phi^2} -\frac{1}{2}\bar{m}^2\Phi^2+\frac{\Lambda}{4}\big(2\pi^2 a^3\big)^{-1}\Phi^4
\end{equation}
where  $\mathbf{j}(\mathbf{j}+2)$  are eigenvalues of angular momentum operators of the internal $SO(4)$ symmetry field space, $\mathbf{J}$.  \cite{SO4jaber, SO4zhao},

In this section, all results for the norm of Higgs fields will be similar to section(\ref{section3}) and section(\ref{section4}). Here, only after replacing  $\mathsf{j}(\mathsf{j}+1)$ with $\mathbf{j}(\mathbf{j}+2)$,  we can obtain all expressions of the norm of Higgs fields derived as in previous sections.  Moreover,  we can obtain similar results for not only the internal SO(4) symmetric scalar field space,  but also scalar fields belonging to a representation of a compact semi-simple Lie group.

\section{Conclusion}

We propose a model in which the size of the universe is related to Higgs mass. In contrast to the Standard Model, we have worked in FLRW space-time.  Thanks to the symmetries of the closed FLRW universe, the size of the universe can be included in our calculations (\ref{com}). After the first quantization of the scalar field background, spontaneous symmetry breaking has been obtained by using the properties of SO(3) symmetry in the internal scalar field space. By taking the minimum of the effective potential, the Higgs field's dependence on the size of the universe can be calculated. The general solution is too complicated to get an idea. We did some assumptions to see how the Higgs field and the size of the universe are related for various simple cases. Moreover, we found that the Higgs field is inversely proportional to the size of universe by some power of scale sizes: $a^{-1}$ (\ref{eqnhiggs1}) and $a^{-\frac{3}{2}}$ (\ref{eqnhiggs2}). Moreover, in FLRW cosmology, the scale factor  describes the expansion of the universe over time.  The character of the scale factor changes according to different eras: matter-dominated era, dark-energy-dominated era and radiation-dominated era. Thus, we have been able to show the behaviour of the Higgs field for different periods of the universe. Finally, instead of the SO(3) internal symmetric scalar field space, we have started with the Higgs field represented by the complex doublet constructed by SO(4) symmetric four scalar fields, as in the Standard Model. By following the same procedure, similar results have been obtained. We are investigating models where the Higgs field arises more realistically as an essential future of this model.

\bibliography{references}

\begin{thebibliography}{9}


\medskip


\bibitem{wald}
 R.M. Wald 
\textit{''General Relativity''}
University of Chicago Press (edition UK ed.), (1984)


\bibitem{rovelli}
C. Rovelli
\textit{''General Relativity: The most beautiful of theories''}
Volume 28 in the series De Gruyter Studies in Mathematical Physics (2015)

\bibitem{AEinstein}
A. Einstein
\textit{''Explanation of the Perihelion Motion of Mercury from the General Theory of Relativity''}
Sitzungsber.Preuss.Akad.Wiss.Berlin (Math.Phys.) 831-839 (1915) 


\bibitem{preccession}
R.S. Park, W.M. Folkner, A.S. Konopliv, J.G. Williams, D.E. Smith, and M.T. Zuber
\textit{''Precession of Mercury's Perihelion from Ranging to the MESSENGER Spacecraft''}
The Astronomical Journal, Volume 153, Number 3


\bibitem{blackhole}
R. M. Wald
\textit{''Gravitational Collapse and Cosmic Censorship''}
Black Holes, Gravitational Radiation and the Universe. Dordrecht: Springer. pp. 69–86. arXiv:gr-qc/9710068. doi:10.1007/978-94-017-0934-7


\bibitem{bhole}
M. F. Wondrak, W.D. van Suijlekom, H. Falcke
\textit{''Gravitational Pair Production and Black Hole Evaporation''}
Phys. Rev. Lett. 130 (2023) 221502


\bibitem{lensing}
M. Bartelmann
\textit{''Gravitational lensing''}
Classical and Quantum Gravity, Volume 27, Number 23


\bibitem{ligo}
 Abbott et al. 
\textit{''Observation of Gravitational Waves from a Binary Black Hole Merger''}
Phys. Rev. Lett. 116, 


\bibitem{wave}
F.R. Ares, O. Henriksson, M. Hindmarsh, Carlos Hoyos, and Niko Jokela
\textit{''Gravitational Waves at Strong Coupling from an Effective Action''}
Phys. Rev. Lett. 128, 131101  29 March 2022


\bibitem{originuniverse}
M. Tanabashi et al. (Particle Data Group)
\textit{''Review of Particle Physics''}
 p. 358, chpt. 21.4.1: "Big-Bang Cosmology" (Revised September 2017) by Keith A. Olive and John A. Peacock.Phys. Rev. D 98, 030001  (2018)





\bibitem{einsteinequation}
A.D. Rendall
\textit{''Theorems on Existence and Global Dynamics for the Einstein Equations''}
Logo of springeropen
Living Reviews in Relativity
Living Rev Relativ. (2005)


\bibitem{einsteinfieldequations}
J.K. Singh , R. Nagpal
\textit{''FLRW cosmology with EDSFD parametrization''}
 Eur. Phys. J. C 80, 295 (2020)

\bibitem{flrwhom}
I.I. Cotăescu
\textit{''Dynamical particles in spatially flat FLRW space-times''}
 Eur. Phys. J. C volume 82, Article number: 86 (2022)


\bibitem{flrwstructure}
T. Wijenayake and M. Ishak
\textit{''Expansion and growth of structure observables in a macroscopic gravity averaged universe''}
Phys. Rev. D 91, 063534 (2015) 30 March 2015



\bibitem{fqflrw}
N. Dimakis, A. Paliathanasis, M. Roumeliotis, and T. Christodoulakis
\textit{''FLRW solutions in f(Q) theory: the effect of using different connections''}
Phys. Rev. D 106, 043509  (2022)



\bibitem{Standard}
J. M. Butterworth
\textit{''The Standard Model: how far can it go and how can we tell?''}
Phil. Trans. R. Soc. A.3742015026020150260


\bibitem{electroweak}
D. Britzger, M. Klein, H. Spiesberger
\textit{''Electroweak Physics in Inclusive Deep Inelastic Scattering at the LHeC''}
Eur. Phys. J. C (2020) 80: 831



\bibitem{qcd}
K.J. Eskola, P. Paakkinen, H. Paukkunen, C.A. Salgado
\textit{''EPPS21: a global QCD analysis of nuclear PDFs''}
Eur. Phys. J. C (2022) 82: 413



\bibitem{atlas}
G. Aad, T. Abajyan, B. Abbott, J. Abdallah,S.A.Khalek,A ,et al.
\textit{''Observation of a new particle in the search for the Standard Model Higgs boson with the ATLAS detector at the LHC''}
Physics Letters B
Volume 716, Issue 1, 17 September 2012, Pages 1-29

\bibitem{cms}
S.Chatrchyan, V. Khachatryan, A. M.Sirunyan, A.Tumasyan,W. Adam, E. Aguilo, T. Bergauer, et al.
\textit{''Observation of a new boson at a mass of 125 GeV with the CMS experiment at the LHC''}
Physics Letters B
Volume 716, Issue 1, 17 September 2012, Pages 30-61


\bibitem{kondo}
K. Kondo
\textit{''Gauge-independent Brout–Englert–Higgs mechanism and Yang–Mills theory with a gauge-invariant gluon mass term''}
 Eur. Phys. J. C volume 78, Article number: 577 (2018)



\bibitem{standardmodel11}
C.T. Hill, E.H. Simmons
\textit{''Strong dynamics and electroweak symmetry breaking''}
Physics Reports
Volume 381, Issues 4–6, July 2003, Pages 235-402



\bibitem{higgsboson}
N. Becerici Schmidt, S.A. Çetin, S. Iştin,  S. Sultansoy 
\textit{''The fourth Standard Model family and the competition in Standard Model Higgs boson search at Tevatron and LHC
''}
 Eur. Phys. J. C volume 66, pages119–126 (2010)



\bibitem{diameter}
I. Bars, J.
 Terning
\textit{''Extra Dimensions in Space and Time''}
 Springer. pp. 27 (2009)

\bibitem{peskin}
M.E. Peskin , D.V. Schroeder 
\textit{''An Introduction To Quantum Field Theory''}
1995, Addison-Wesley


\bibitem{higgsflrw}
D.P. George, S. Mooij and M. Postma
\textit{''Effective action for the Abelian Higgs model in FLRW''}
Journal of Cosmology and Astroparticle Physics, Volume (2012)

\bibitem{Arttu}
A. Rajantie
\textit{''
Higgs cosmology''}
Phil. Trans. R. Soc. A.376, (2018)



\bibitem{shapos}
M. Shaposhnikov
\textit{''The Higgs boson and cosmology''}
Phil. Trans. R. Soc. A.3732014003820140038 (2015)


\bibitem{supergravity}
Y. Aldabergenov, S.V. Ketov
\textit{''Higgs mechanism and cosmological constant in N=1 supergravity with inflaton in a vector multiplet''}
The European Physical Journal C volume 77, Article number: 233 (2017) 


\bibitem{cosmologicalcons}
I. Dymnikova
\textit{''The Higgs Mechanism and Cosmological Constant Today''}
Universe, vol. 8, issue 6, p. 305  (2022)


\bibitem{phase9}
B. Bergerhoff, C. Wetterich 
\textit{''Electroweak Phase Transition in the Early Universe?''}
Current Topics in Astrofundamental Physics: Primordial Cosmology pp 211–240
the NATO ASI Series book series (ASIC,volume 511) (1996)




\bibitem{phasetransition}
D. Boyanovsky, H.J. de Vega,and D.J. Schwarz4
\textit{''Phase Transitions in the Early and Present Universe''}
Annual Review of Nuclear and Particle Science. 56:441-500 (2006)




\bibitem{eweak}
D.D. Dietrich
\textit{Electroweak symmetry breaking in other terms}
 Eur. Phys. J. C volume 67, pages237–252 (2010)

\bibitem{eweak2}
H.H. Patel and M.J. Ramsey-Musolf
\textit{Stepping into electroweak symmetry breaking: Phase transitions and Higgs phenomenology}
Phys. Rev. D 88, 035013 (2013)


\bibitem{Electroweaksymmetry}
M. Shaposhnikov, A. Shkerin, and S. Zell
\textit{''Standard model meets gravity: Electroweak symmetry breaking and inflation''}
Phys. Rev. D 103, 033006  (2021)

\bibitem{eweak3}
M. Hindmarsh, M. Lüben, J. Lumma, M. Pauly
\textit{''Phase transitions in the early universe''}
SciPost Phys. Lect. Notes 24 (2021)


\bibitem{inflaton}
F. Bezrukov , M. Shaposhnikov 
\textit{''The Standard Model Higgs boson as the inflaton''}
Physics Letters B
Volume 659, Issue 3, 703-706 (2008)


\bibitem{Inflation}
A.O. Barvinsky, A.Y. Kamenshchik and A.A. Starobinsky
\textit{''Inflation scenario via the Standard Model Higgs boson and LHC''}
JCAP11(2008)021


\bibitem{darkenergy}
O. Lebedev
\textit{''The Higgs portal to cosmology''}
 Progress in Particle and Nuclear Physics


\bibitem{darkmatter}
G. Arcadi, A. Djouadi, M. Raidal 
\textit{''Dark Matter through the Higgs portal''}
Physics Reports
Volume 842, 3 February 2020, Pages 1-180




\bibitem{higgsportal1}
G. Arcadi, A. Djouadi  M. Kado 
\textit{''The Higgs-portal for dark matter: effective field theories versus concrete realizations''}
 Eur. Phys. J. C volume 81, Article number: 653 (2021)





\bibitem{frwspace}
S.K. Modak , D. Singleton
\textit{''Baryogenesis via Hawking-like radiation in the FRW space-time''}
 Eur. Phys. J. C volume 75, Article number: 200 (2015) 


\bibitem{frwspace2}
S. Cao, J. Qi, Z. Cao, M. Biesiada, J. Li, Y. Pan, Z. Zhu
\textit{''Direct test of the FLRW metric from strongly lensed gravitational wave observations''}
Scientific Reports volume 9, Article number: 11608 (2019) 




\bibitem{expansion}
V. Vavrycuk
\textit{''Gravitational orbits in the expanding Universe revisited
''}
Front. Astron. Space Sci. (2023)


\bibitem{expansion2}
B. Holdom
\textit{''Cosmologies with turning points
''}
Physics Letters B
Volume 839,  (2023)

\bibitem{calmet}
X. Calmet
\textit{'' Cosmological evolution of the Higgs boson’s vacuum expectation value
''}
Eur. Phys. J. C 77, 729 (2017)


\bibitem{angular1}
J.L. Jaramillo, E Gourgoulhon 
\textit{''Mass and Angular Momentum in General Relativity''}
Mass and Motion in General Relativity pp 87–124
Part of the Fundamental Theories of Physics book series (FTPH,volume 162) (2010)



\bibitem{sakurai}
J.J. Sakurai
\textit{''Modern Quantum Mechanics''}
Revised Edition, Addison-Wesley Publishing Company, Reading, MA. (1994)



\bibitem{grove}
L.C. Grove
\textit{''Classical Groups and Geometric Algebra''}
American Mathematical Society (2002)

\bibitem{isos}
C. Rogers, W.K. Schief
\textit{''Bäcklund and Darboux Transformations:
Geometry and Modern Applications in Soliton Theory''}
cambridge (2002)


\bibitem{angularmomentum}
M. Arik, T. Tok
\textit{Angular momentum in the internal SO(3) symmetry space relating the size of universe and the Higgs mass}
arXiv:2202.12432 (2022)


\bibitem{tachyon}
G. Felder, Lev Kofman, A. Linde
\textit{''Tachyonic instability and dynamics of spontaneous symmetry breaking''}
Phys. Rev. D 64, 123517  (2001)

\bibitem{cosmologicaltime}
J. Magueijo
\textit{''Cosmological time and the constants of nature''}
Physics Letters B
Volume 820, 136487 (2021)

\bibitem{higgstacyonic}
V. Branchina, E. Messina
\textit{''Stability,Higgs Boson Mass and New Physics''}
Phys. Rev. Lett. 111, 241801 – Published 10 December (2013)



\bibitem{Jordan}
O. Dunya, L. Akant, M. Arik, Y. Kardas, S. Sahin  T. Tok 
\textit{''The Higgs field and the Jordan Brans Dicke cosmology''}
 Eur. Phys. J. C volume 81, Article number: 66 (2021)

\bibitem{scalarmode}
M. Arık, T. Tok
\textit{''The scalar mode of gravity''}
Modern Physics Letters AVol. 37, No. 10, 2250068 (2022)









\bibitem{flrwhistory1}
A. Friedmann
\textit{'' Uber die Krummung des Raumes''}
Zeitschrift fur Physik A, 10 (1): 377-386, (1922).

\bibitem{flrwhistory2}
G. Lemaitre
\textit{''A Homogeneous Universe of Constant Mass and Growing Radius Accounting for the Radial Velocity of Extragalactic Nebulae''}
 Annales Soc.Sci.Bruxelles A 47  49-59 (1927), Gen.Rel.Grav. 45 1635-1646 (2013) 


\bibitem{flrwhistory3}
H. P. Robertson
\textit{''Kinematics and world structure''}
Astrophysical Jour- nal, 82: 284-301, 284-301 (1935)

\bibitem{flrwhistory4}
A. G. Walker
\textit{''On Milne’s theory of world-structure''}
Proceedings of the London Mathematical Society, Series 2, 42 (1): 90-127 (1937)


\bibitem{dominatedcosmology}
B.S. Ryden
\textit{'' Introduction to Cosmology''}
Cambridge University Press (2016)



\bibitem{SO4jaber}
S.M. Al-Jaber
\textit{''Quantization of angular momentum in the  
N-dimensional space''}
Nuov Cim B 110, 993–995 (1995)


\bibitem{SO4zhao}
Z. Wei-Qin
\textit{''Relation Between Dimension and Angular Momentum for Radially Symmetric Potential in N-Dimensional Space''}
 Commun. Theor. Phys. 46 429 (2006)





\end{thebibliography}

\end{document}